\documentclass[doublespacing]{elsart}
\usepackage{graphicx}
\usepackage{float}
\usepackage{amsfonts}
\usepackage{natbib}

\begin{document}
\begin{frontmatter}
\title{Crustal structure below Popocat\'epetl Volcano (Mexico) from analysis 
of Rayleigh waves.}
\author[Grenoble]{Louis De Barros\corauthref{cor1}}, 
\ead{louis.debarros@obs.ujf-grenoble.fr}
\author[Grenoble]{Helle A. Pedersen},
\author[Chambery,IRD]{Jean Philippe M\'etaxian},
\author[Mexique] {Carlos Vald\'es-Gonzalez} and
\author[Chambery,IRD,Mexique]{Philippe Lesage}
\address[Grenoble]{Laboratoire de G\'eophysique Interne et Tectonophysique, 
Observatoire des Sciences de l'Univers de Grenoble, BP 53, 38041 Grenoble 
Cedex 9, France.}
\address[Chambery]{Laboratoire de G\'eophysique Interne et Tectonophysique,
Universit\'e de Savoie, 73376 Le Bourget-du-Lac Cedex, France.}
\address[IRD]{Institut de Recherche pour le D\'eveloppement, France.} 
\address[Mexique]{Instituto de Geof\'\i sica, 
Universidad Nacional Autonoma de M\'exico, Ciudad Universitaria, Del. Coyoacan, 
M\'exico D.F., CP 04510 M\'exico.}
\corauth[cor1]{Corresponding author:} 


\begin{abstract}
An array of ten broadband stations was installed on the Popocat\'epetl volcano 
(Mexico) for five months between October 2002 
and February 2003. 26 regional and teleseismic earthquakes were selected and filtered
in the frequency time domain to extract the fundamental mode of the Rayleigh 
wave. The average dispersion curve was obtained in two steps.
Firstly, phase velocities were measured in the period range [2 - 50] s 
from the phase difference between pairs of stations,
using Wiener filtering. Secondly, the average dispersion curve
was calculated by combining observations from all events in order to reduce 
diffraction effects. The inversion of the mean phase velocity yielded 
a crustal model for the volcano which is consistent
with previous models of the Mexican Volcanic Belt. 
The overall crustal structure beneath Popocat\'epetl is therefore 
not different  from the surrounding area and the velocities in the lower crust
are confirmed to be relatively low. 
Lateral variations of the structure were also investigated
by dividing the network into four parts and by applying the same procedure 
to each sub-array. No well defined anomalies appeared for the two sub-arrays 
for which it was possible to measure a dispersion curve. 
However, dispersion curves associated with individual events reveal important 
diffraction for 6 s to 12 s periods which could correspond to strong 
lateral variations at 5 to 10 km depth.

\end{abstract}
\end{frontmatter}


\section{Introduction}
Popocat\'epetl is a large andesitic strato-volcano,
located 60 km south-east of
Mexico City and 40 km West of Puebla (fig. 1.a). It belongs to the 
Trans-Mexican Volcanic
Belt (MVB). Its large cone is the second highest summit of Mexico (5452 m 
above sea level) with an elipsoidal 600-800 m wide crater.\\
The present active period began on December 21st 1994.
Since 1996, an andesitic to  dacitic dome cyclicly grows into the crater 
and bursts producing high plumes of gas and ash \citep{arcie99, wright02}. 
More than 100 000 persons could potentially be directly affected by an 
eruption and ashes could affect an area with more than 20 million people 
\citep{cruzreyna97, macias05}.\\
The overall crustal structure beneath the MVB is relatively well studied
\citep{campil96, valdes86, shapiro97}.
On the contrary, the crustal seismic structure beneath Popocatepetl is not 
well known. Receiver functions analysis by \citet{cruzatienza01},
using 4 events from South America, indicates that a Low Velocity Zone
may be present beneath a station located 5 km north of the crater.\\ 
The aim of this paper is to improve the knowledge
of this complex volcano structure, 
and particularly to determine
if the whole crust beneath the volcano is
significantly different from the rest of the MVB. The first kilometers
of crust beneath several volcanoes have been studied 
 \citep[e.s.][]{Dawson, Laigle, Benz}. Typical volcanic 
anomalies are low velocity zones, attributed to the presence of partial melt, 
or high velocity zones, due to solidified magmatic intrusions.\\
We concentrate on the S-wave structure, as S-wave velocities are
very sensitive to temperature changes and to the presence of even small 
amounts of partial melt. The easiest way to get an overall 
picture of S-wave velocities is through surface wave analysis. 
However, the traditional 2-stations methods 
can not be used in this rather diffractive environnement as measurements would 
possibly be strongly biased  due to local and regional diffraction 
\citep{Wielandt, Friederich}. An alternative approach is therefore 
to use array analysis. Such methods have been used on volcanoes, particularly 
for tremor source location  \citep{Metaxian, Almendros} or for 
shallow structure study \citep[See for example][]{Saccorotti}. 
More details can be found in \citet{chouet} who presents a state of the art 
on volcano seismology.\\
The dispersion curve had to be measured over a wide 
frequency range (0.02-1 Hz) to study the overall crustal structure beneath 
Popocat\'epetl. The array configuration which was strongly influenced by 
topography and logistic issues was such that we could not use
spatial Fourier transforms outside a very narrow frequency range.
Consequently methods based on wavenumber decomposition were excluded. 
The use of time-domain methods was problematic as we needed 
a good frequency resolution.\\ 
These considerations led us to use the procedure of \citet{peder03} to 
measure phase velocities across the array.
The assumption behind this method is that the records are constituted 
by one single plane wave which propagates through the array. 
 Even though this hypothesis is most probably wrong for most individual events,
it may be corrected by averaging out unwanted waves (diffraction effects, 
non plane waves, etc) using events from different directions.  
The variability between different events 
will also provide an error estimate on the dispersion curve.
To increase frequency range and azimuthal coverage we used
both teleseismic and local events.\\
After a short description of the data and the processing methods used, 
we present and discuss the main results, with a comparison of 
the overall crustal structure beneath the volcano to that of the MVB.\\
\section{Data}
 An array of nine stations (Guralp CMG 40T)
with three-components broadband sensors (30-60 s cut-off period) 
was installed in October 2002 on the Popocat\'epetl 
volcano and continuously recorded
four months of seismic events. Figure 1.b shows the array geometry.
The station altitudes were between 2500 and 4300 m above sea level. 
The reference altitude used in this 
study corresponds to an average level of 3500 m a.s.l.\\  
To obtain dispersion curves in a period range of 2 to 50 s, we chose to use 
both teleseimic and local events with epicentral distances
between 200 and 15000 km (see fig. 2). We selected vertical components of 
events with a good signal to noise ratio and with well developed Rayleigh 
waves.  The usuable frequency range for the two types of events overlapped, 
however the long period part of the dispersion curve was mainly calculated 
using teleseismic events while the shorter periods were dominated by regional 
events.\\  
Prior to the array analysis, we deconvolved the data with the instrument 
responses.  The second step of this analysis was to enhance the 
signal-to-noise ratio through
 time-frequency filtering \citep{levshin89}. In this part of the analysis we 
firstly applied multiple filter to the data and identified the group velocity 
dispersion curve by the maximum amplitude at each frequency. We secondly 
integrated this curve to obtain the phase velocities and subtracted the 
corresponding phase $\theta(f)$  at each frequency to obtain a non dispersive 
signal. 
A time window was then applied onto the non-dispersive wave to suppress noise 
and the phase $\theta(f)$ was finally added. Fig. 3 shows the comparison 
between an unfiltered record (3.a), with its corresponding group 
velocity (3.c), and filtered record (3.b) of a teleseismic event. \\
The time-frequency filter efficiently reduces the influence of noise, 
body waves and higher mode Rayleigh waves. It also makes it possible to 
identify and exclude events whose fundamental mode of Rayleigh waves is not 
well separated from other waves. 
10 events were rejected during this stage. A further 12 events were excluded 
during  the array analysis, leaving 26 events.
Table 1 contains the final event list, and figure 2 shows their
distribution. The number of teleseismic events was too small to ensure
a correct back-azimuth distribution, but there were events from all quadrants.
The regional events were mainly located in the Pacific Coast subduction
and the Caribbean Islands, ensuring a back-azimuth range between N126$^o$ 
(South 
East) to N273$^o$ (West).
\section{Methodology}
To measure phase velocities, we follow \citet{peder03}.
In this method, the phase velocity at a given frequency is obtained 
in two steps. Firstly, each event is analysed independently. 
 In this step, the phase $\phi$ of the Wiener filtering $W(f)$ is transformed 
into  time delays $\Delta t$ between each pair of stations using 
$\Delta t=\phi /( 2 \pi f)$.
The Wiener filtering in the frequency domain that we use is given by:
\begin{displaymath}
W(f)=\frac{S_{XY} \ast Han(f) e^{j 2 \pi f t_0}}{\sqrt{ S_{XX} \ast Han(f)}\sqrt{ S_{YY} \ast Han(f)}}
\end{displaymath}
$S_{XX}$, $S_{YY}$ and $S_{XY}$ are respectively the Fourier transforms of the 
autocorrelations and intercorrelation of the two signals , $Han(f)$ is the 
Fourier transform of the Hanning fonction $han(t)$ and $t_0$ is the delay of 
the intercorrelation peak. $\ast$ represents the convolution product. 
A grid-search  on velocity and back-azimuth is applied to find the best 
fitting plane wave that would explain the observed time delays. The fit is 
calculated with the L1 norm, i.e the average absolute difference between 
observed and predicted time delays. The knowledge of the back-azimuth 
makes it possible to subsequently calculate the distance between each pair
of stations projected onto the slowness vector. In this way each event yields 
a series of (distance,delay) points.  \\
Secondly, a bootstrap
process is applied \citep{Schorlemmer,Efron}:
500 bootstrap samples are created  by resampling the 26 events
of the data set. For each bootstrap sample, the phase velocity is calculated
as the inverse of the slope of the best fitting line through all 
the (distance, delay) points.  The L1 norm is also used here to estimate the 
fit between observations and predictions. The points are associated with 
weighting which reflects how well the data fitted the assumption of a plane 
wave in the first step of the analysis.  The final phase velocity and the 
associated
uncertainity are obtained as the average and standard deviation over
the 500 samples.\\
The advantages of this method are 1) stabilization of delay measurements
through Wiener filtering; 2) stabilization of back-azimuths
and phase velocities through the use of the L1 norm;
3) weighting of the events in the final phase velocity calculation
according to the quality of the back-azimuth estimate;
4) estimation of realistic error bars on the final dispersion curve.
For more details, we refer to \citet{peder03}.\\
The last part of this analysis consists in inverting the dispersion curves.
We used the two-step inversion methods proposed by \citet{shapiro97}.
Firstly, the average dispersion curve was inverted using a linearized,
classical dispersion scheme \citep{herrmann87} to find a simple
shear wave model which fitted the dispersion curve.
We then used this model for a stochastic nonlinear Monte-Carlo inversion.
Interface depths and shear-wave velocities were randomly changed into a new 
model which was kept and used in the next iteration if it fitted the dispersion
curve within the error bars. This second step was repeated 5000 times.
 We eliminated unrealistic models by applying loose constraints 
on Moho depth (between 40 and 50 km depth) and on the S-wave velocity near the 
surface in agreement to the existing models (between 1.5 and 3 km/s).  
We finally calculated the average model and we verified that the corresponding 
dispersion curve fitted within the error bars of the observed one.  The error 
bars of the final models were computed as the standard deviation of all the 
acceptable models.   
Quality factors and  P-wave velocities were kept constant during the inversion 
as their influence was significantly smaller than the error bars 
of the dispersion curve. 
\section{Results}
To detect differences between the crust under the Popocat\'epetl and the 
standard crust of the MVB, one can compare the equivalent shear wave 
velocities profiles. As surface wave inversions are non-unique it is however 
useful to also compare the dispersion curves which correspond to the 
existing models.\\ 
The models that we compare with were from: 
1) \citet{cruzatienza01} who obtained their model through inversion of 
receiver functions using four teleseismic events from South America at 
station PPIG (located 5 km north of the Popocat\'epetl crater, see fig.1); 
2) \citet{campil96} who inversed the group velocities of local events between 
the Guerrero Coast and Mexico City; 3) \citet{valdes86} whose model is the 
result 
of a seismic refraction study in Oaxaca. We recalculated the phase velocities 
corresponding to these models. \citet{shapiro97} detected lateral variations
of uppermost crustal structure within the MVB using surface wave group 
velocities. Due to the limited depth penetration in their study (10 km), 
their models are 
not included in our figures, but will be integrated in the discussion of 
the results.
\subsection{Full array}
We firstly used all the stations and the 26 events to measure the 'overall 
dispersion curve', i.e. the average dispersion curve within the full array. 
The phase velocities were unstable above 35 s period and were not used
in the inversions.\\
In Figure 4.a we compare our dispersion curve with the ones corresponding to 
the earth model derived by \citet{campil96}, \citet{cruzatienza01} 
and \citet{valdes86}. For periods longer than 8 s, the \citet{campil96}
curve is similar to ours. For short periods, the velocities  
increase rapidly with period, similarly to the \citet{cruzatienza01} curve.\\
We verified that our inversions of the observed phase velocities were 
independent of which of the three reference models (see fig 4.b)  was used as 
starting model.
The results shown here are obtained by using the one of \citet{campil96} as
it has the advantage of fitting our data well and it only has 
four layers. The latter is important to allow for an efficient exploration 
of the parameter space in the Monte Carlo inversion.\\
Our preferred model (fig. 4.b) shows low shear velocities (2.2 km/s) 
between the surface and 3 km depth, overlying a layer with velocities 
increasing slowly from 3.4 to 3.7 km/s between 6 to 20 km depth. 
The transition between the two layers may be either a strong gradient or a 
sharp interface. The lower part of the crust has a constant velocity of 3.75 
km/s down to Moho which is located at 45 km depth. The velocity below Moho 
is approximatively 4.3 km/s. The lower crust and upper mantle velocity as well 
as the Moho depth are not well resolved due to trade-off between 
these parameters and because of a maximal period of 35s.\\
The boundary depths that we obtained are however consistent 
with existing models,
in particular with \citet{campil96}.
Our near surface velocities are however significantly lower and 
our upper crustal velocities slightly higher than those of \citet{campil96}, 
while the two models are virtually almost 
identical in the lower crust. 
 The low velocity of the surface layer is relatively well constrained, 
however we can not exclude that the layer would be slightly thinner with 
slightly lower velocity.  This layer, also identified by \citet{cruzatienza01}, can be associated with the poorly consolidated 
materials of the volcano cone. \citet{shapiro97} find  that the velocities 
in the upper 2 km are low beneath the southern part of the MVB where the 
volcanic activity is recent as compared to the northern part.
Our results imply that the overall crustal structure below Popocat\'epetl 
is not significantly different from that of the MVB.  \\
 We verified whether an 6-layers initial model with a Low velocity zone 
between 6 and 10 km inspired by the \citet{cruzatienza01} model, would yield 
a significantly different result. The resulting model is not different 
from our prefered model (fig 4.b), in particular there is no significant 
Low Velocity Zone. We do not see any indication that the  Low Velocity Zone 
observed by \citet{cruzatienza01} is a general feature of 
the volcano.  
\subsection{Sub-arrays}
To investigate lateral variations within the area, we divided the array into 
sub-arrays for which we calculated dispersion curves independently.  The use 
of sub-arrays was particularly difficult as these arrays were composed of 
only three stations, so technical problems at any of the relevant stations 
would render the analysis impossible. 
It was possible to measure dispersion curves for the Southern 
(South sub-array: FPC, FPP, FPX) and the western sub-array 
(West sub-array: FPA, FPP, FPX).
The dispersion curves for these two sub-arrays are shown in figure 5.  
At the largest period the analysis is mainly based on teleseismic events, 
out of which only 3 or 4 events were avalaible for the sub-array analysis. 
The phase velocity error bars are consequently large at long period 
($>$ 25 s for the South sub-array and $>$ 15 s for the West sub-array).   \\
The dispersion curves for the South sub-array (
located around the active crater) and the West sub-array was respectively
measured with  13 and 14 events (see table 1). 
For the West sub-array, individual dispersion curves show strong oscillations 
between 
6 and 12 s, particularly for events coming from the South or the East.
These oscillations, probably due to local diffraction result in large error 
bar for the final dispersion curve. 
However, the dispersion curves beneath the two sub-arrays are not 
significantly 
different from the overall dispersion curve.\\
 The phase velocities obtained with events for which the surface waves 
propagated through the volcano before encountering the array are more 
fluctuating than those obtained with other events. 
As the majority of the events are located South and South-West of the array, 
the individual 
dispersion curves are more fluctuant with the period at the North and East 
sub-arrays (composed of stations FPA, FPP, FMI and FPC) than for the other 
sub-arrays. It was consequently not possible to calculate a stable dispersion 
curve 
for these two arrays.  \\
The starting model for the inversion for the sub-arrays South and West is the 
model found with the full array, approximated by four layers. 
The estimated velocities are 
not significantly different from those of the overall model. Nevertheless, 
for the South array velocities are slightly smaller between 6 and 10 km depth, 
and the surface velocities are higher. The differences are however smaller 
than the error bars of the overall model.
\section{Discussion and Conclusion}  
The average crust beneath the Popocat\'epetl volcano appears to be similar to
the MVB crust. There is therefore no indication of large scale
crustal anomalies associated with  Popocat\'epetl as compared to the MVB.
We do however confirm that the lower crust in the area is likely to be 
associated 
with relatively low shear wave velocities (3.75 km/s). Close to the surface, 
the velocity is approximatively 2.2 km/s over at least a depth of 3 km.
It probably corresponds to the poorly consolidated material of the cone (such 
as volcanic slags and ash and pyroclastic deposits) overlying the 2 km-thick 
volcanic layer of the MVB crust \citep{shapiro97}.\\
  We speculate that the oscillations observed for the sub-arrays between 
6 and 12 s periods is associated with diffraction by lateral heterogeneity 
at 5-10 km depth as this period range corresponds to wavelengths between 16 
and 36 km. To obtain strong diffraction, the heterogeneity must be of 
considerable size (i.e. of the order of the wavelength), as surface waves 
are not strongly diffracted by many small heterogeneities \citep{chammas03}. 
However, the lack of Low Velocity Zone turns down the hypothesis of a large 
continuous magma chamber. We speculate that either the interface located at 
4 km depth in the average model fluctuate strongly, or that an abrupt lateral 
change takes place immediately beneath the central part of the volcano.  
The unresolved velocities at 5-10s 
period at the West and South sub-arrays indicate that future arrays should be 
designed so as to give good constraints at 5-10 km depth. To obtain this, 
more stations and a larger recording period are necessary.\\
 More seismic events with a better 
azimutal distribution would improve the smoothing and the error bars of the 
dispersion curves and make it possible to include receiver function analysis 
and coupled Rayleigh-Love inversion. \\
\begin{ack}
We are grateful to Germ\'an Espitia-Sanchez of the Centro Nacional de 
Prevencion de los Deasatres (CENAPRED), Marcos Galicia-L\'opez of the 
Instituto de Protecci\'on Civil del Estado de M\'exico, Sargento Fidel Limon 
of the VI Region Militar, Aida Quezada-Reyes and Ra\'ul Ar\'ambula-Mendoza of 
the Universidad Nacional Aut\'onoma de M\'exico for their logistical support 
and participation in the field experiment. We also thank the local department 
in M\'exico of 
Institut de Recherche pour le D\'eveloppement for its support during 
the field work. Funding for the experiment was provided by the Centre National 
de la Recherche Scientifique (PNRN-INSU 2000 and 2002), the Coordination de la 
Recherche Volcanologique, the Universit\'e de Savoie (BQR 2002), the ARIEL 
program and by the CONACYT project 41308-F. Most of the seismic stations were 
provided by the R\'eseau Acc\'el\'erom\'etrique Mobile (RAM-INSU).H. A. 
Pedersen received financial support from the Alexander von Humboldt Foundation. Servando de la Cruz-Reyna and an anonymous reviewers made valuable comments to improve the manuscript.
\\
\end{ack}

\newpage

\textbf{\Large{TABLE AND FIGURE CAPTIONS}}

Table 1: Origine time, epicentre distance and back-azimuth of the events used 
for measuring the dispersion curves for the full array or with the sub-arrays 
(columns 5 and 6).\\

Figure 1: (a) Location of the Popocat\'epetl volcano and (b)
 array geometry used in the analysis. PPIG is a permanent 
station used by 
\citet{cruzatienza01} and is not used in this study.\\

Figure 2: Location and azimuth distribution of (a) teleseismic and (b) local events used in the array analysis.\\

Figure 3: Example of frequency-time filtering:
a) Trace recorded at FPX, of the event at 03:37:42 GMT 
on november 03th 2002;
b) Same trace after filtering;
c) Group velocity of this event before filtering.\\

Figure 4:  a: Comparison of dispersion curves:\protect\\
1) Uncertainities of our observed dispersion curves ($\pm$ 1 standard deviation, grey area) for the full array and 
2) dispersion curve calculating with our average model (solid line);
3) Dispersion curve for \citet{valdes86};
4) Same for \citet{campil96};
5) Same for \citet{cruzatienza01}.\protect\\
Fig. 4. b: Comparison of crustal models:\protect\\
1) Error bars ($\pm$ 1 standard deviation, grey area) and 2) average S-wave velocity model (solid line); 3) Crustal model for \citet{valdes86};
4) Same for \citet{campil96};
5) Same for \citet{cruzatienza01}.\\

Figure 5: Dispersion curves of the full array (dotted line) and the two sub-arrays (solid line) with their uncertainities (grey area):
a) South sub-array;
b) West sub-array. 
In insert: sub-array geometry and back-azimuths of the events used.

\newpage

\begin{table}[h]
\renewcommand{\arraystretch}{0.5}
\begin{tabular}{cccccccc}
\hline
date & hour & epicentre & back- & West & South \\
 & & (km) & azimuth & array & array \\
\hline
\protect{\vspace{-7pt}}\\
2002/11/03 & 03:37:42  & 11021& 316  & & \\
2002/11/04 & 03:19:18  & 14875 & 79  & &X\\
2002/11/04 & 10:00:47  & 342   & 240 &X&X\\ 
2002/11/04 & 13:57:32  & 366   & 241 &X&X\\
2002/11/05 & 14:05:07  & 650   & 273 &X&X\\
2002/11/06 & 16:02:37  & 279   & 186 &X&X\\
2002/11/06 & 16:24:17  & 321   & 192 &X&X\\
2002/11/06 & 18:04:05  & 390   & 234 &X&X\\
2002/11/07 & 15:14:06  & 7834  & 319 &X&X\\
2002/11/08 & 23:20:41  & 301   & 168 &X&X\\
2002/11/09 & 00:14:18  & 980   & 126 &X&X\\
2002/11/09 & 06:05:58  & 7779  & 164 & & \\
2002/11/15 & 19:58:31  & 10135 & 150 & & \\
2002/11/20 & 22:59:14  & 383  &  233 &X&X\\
2002/11/21 & 02:53:14  & 1903  & 111 &X&X\\
2002/11/26 & 00:48:15  & 7337  & 319 & & \\
2002/11/26 & 16:30:59  & 379  & 236 &X&X\\
2002/11/27 & 01:35:06  & 6511  &323  &X& \\
2002/12/01 & 02:27:55  & 10444 & 234 &X& \\
2002/12/14 & 01:37:48  & 304   & 234 & & \\
2002/12/21 & 08:01:31  & 276   & 189 & & \\
2003/01/21 & 02:46:47  & 1024  & 125 & & \\
2003/01/22 & 19:41:38  & 607   & 268 & & \\
2003/01/22 & 20:15:34  & 618  & 267  & & \\
2003/01/31 & 15:56:52  & 275   & 216 & & \\
2003/02/19 & 03:32:36  & 6749  & 321 & & \\
\hline\\
\end{tabular}
\center{\caption{}}
\label{table1}
\end{table}

\newpage

\begin{figure}[htb]
  \centering
  \includegraphics[height=10cm]{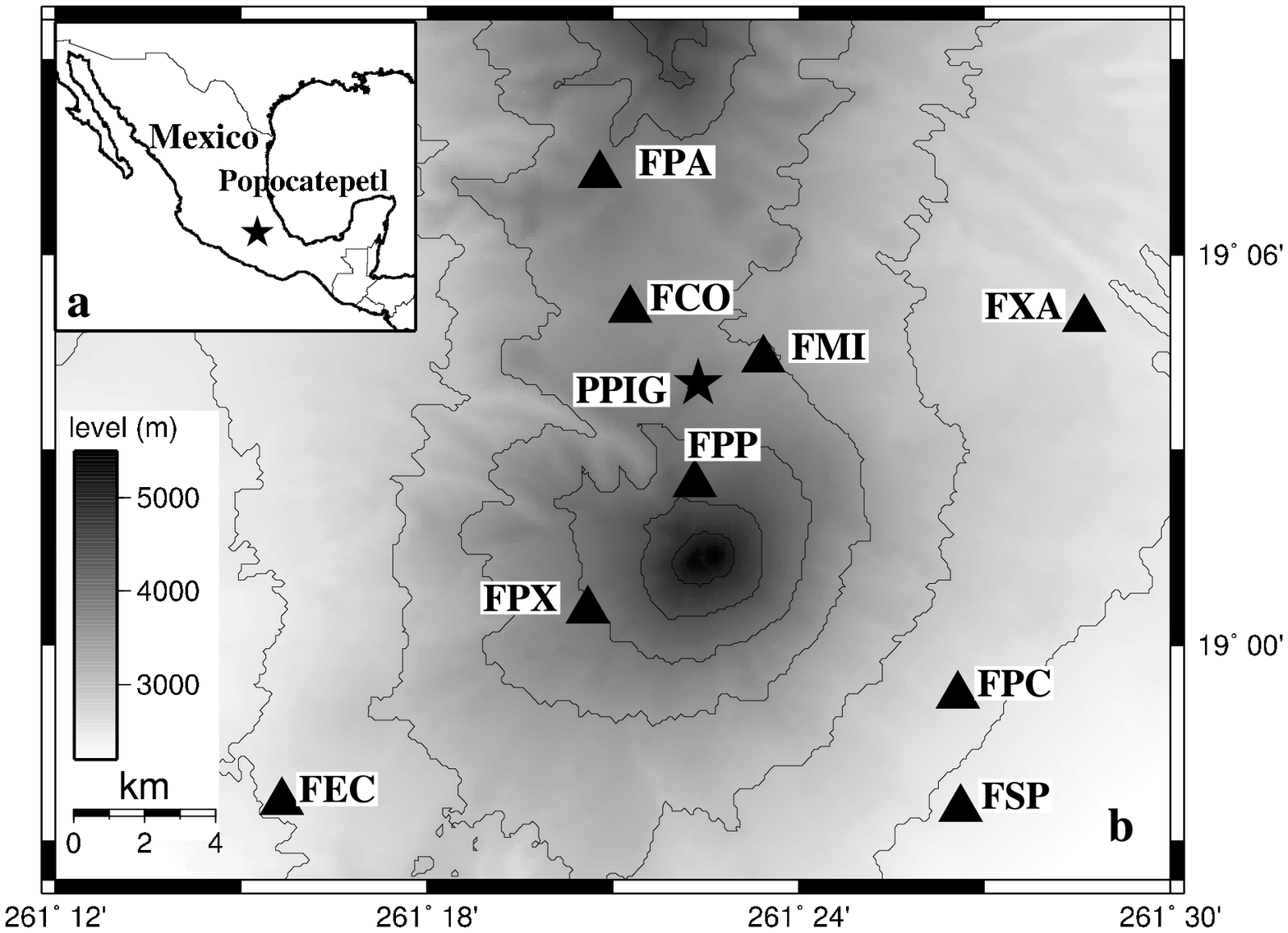}
  \caption{}
 \label{fig1}
\end{figure}

\newpage

\begin{figure}[htb]
  \includegraphics[height=6cm]{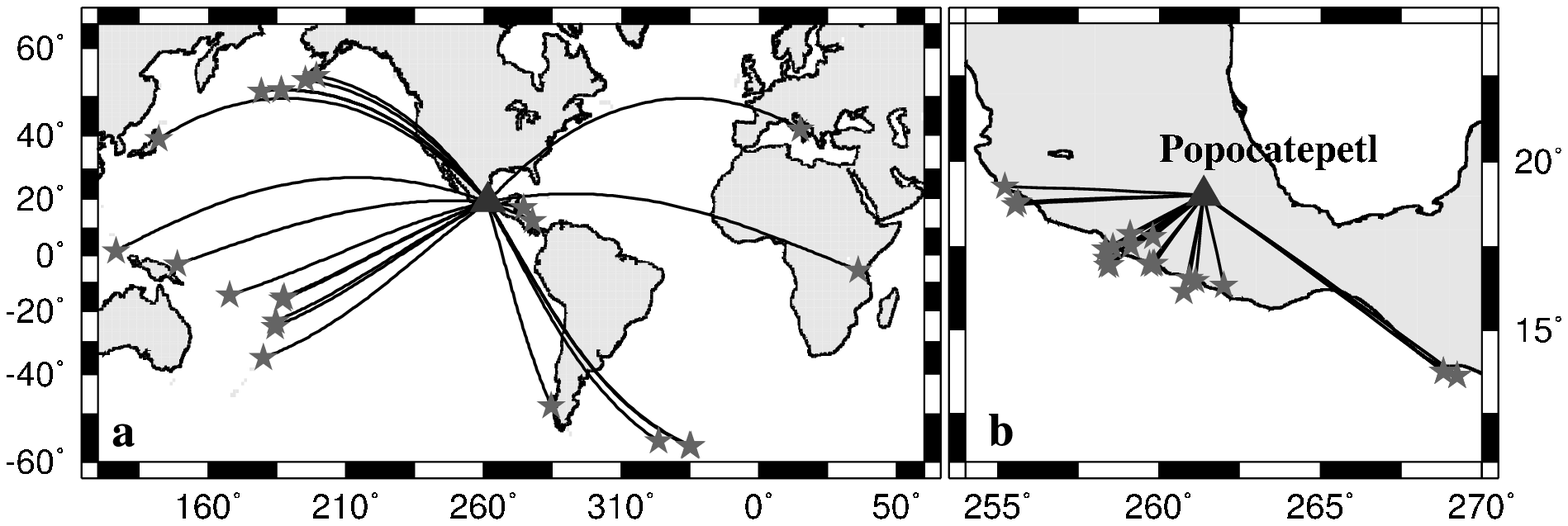} 
\caption{}
 \label{fig2}
\end{figure}

\newpage

\begin{figure}[htb]
  \centering
  \includegraphics[height=7cm]{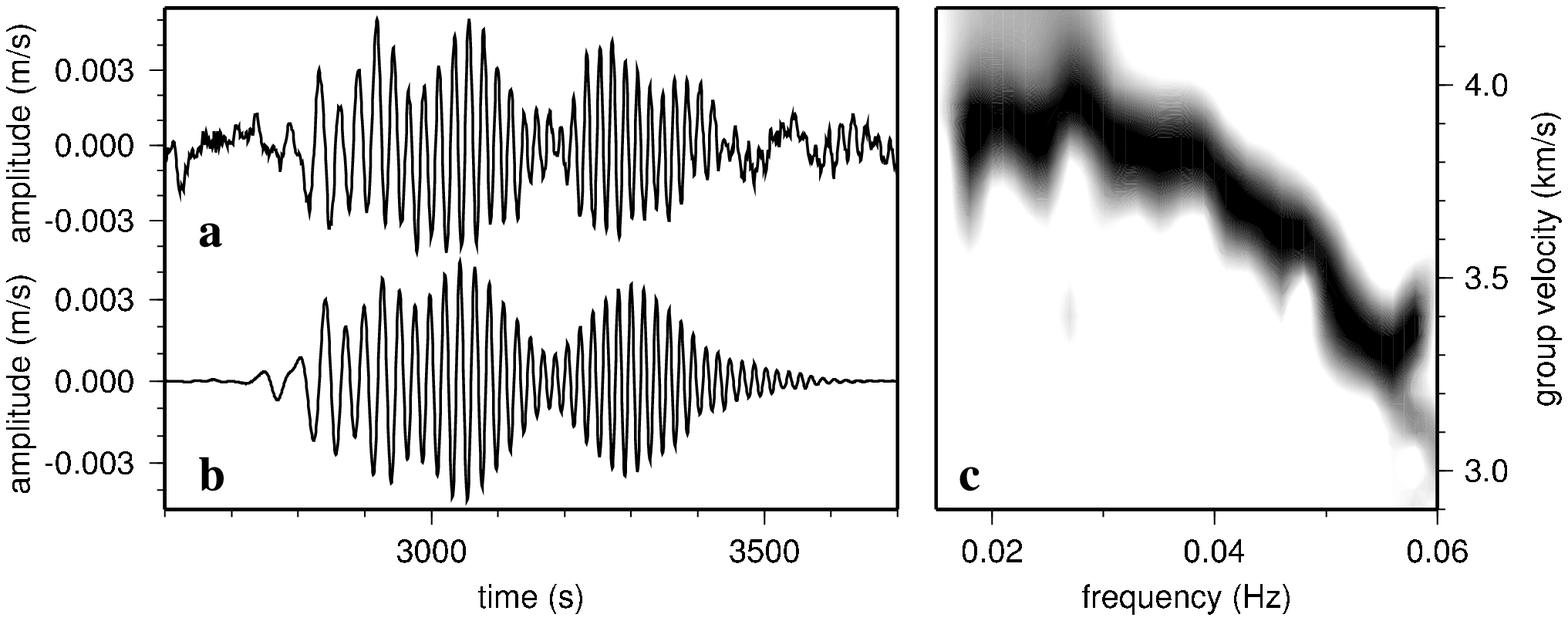}
\caption{}
 \label{fig3}
\end{figure}

\newpage

 \begin{figure}[htb]
  \includegraphics[height=10cm]{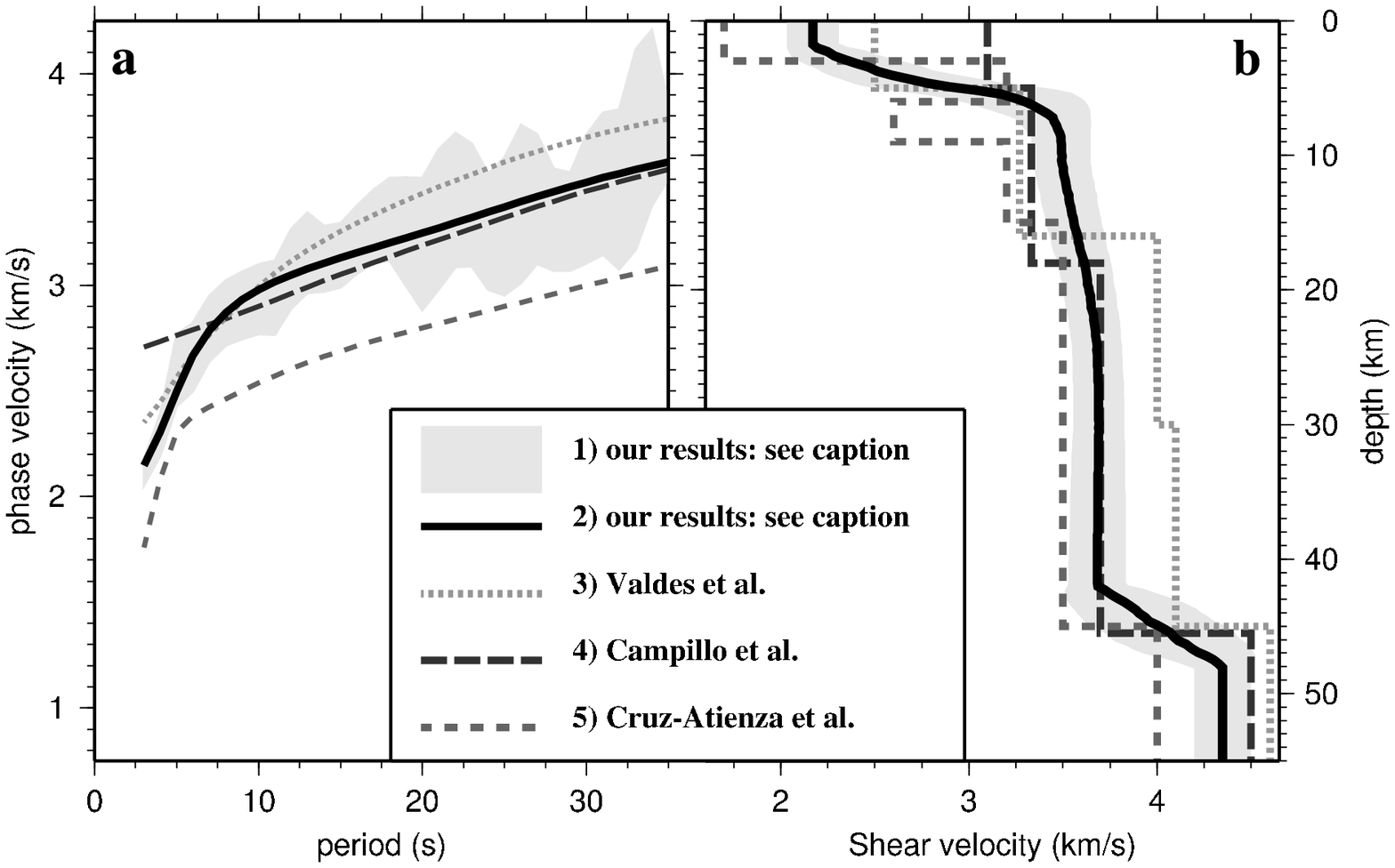}
  \caption{}

 \label{fig4}
\end{figure}

\newpage

 \begin{figure}[htb]
  \centering
  \includegraphics[height=7cm]{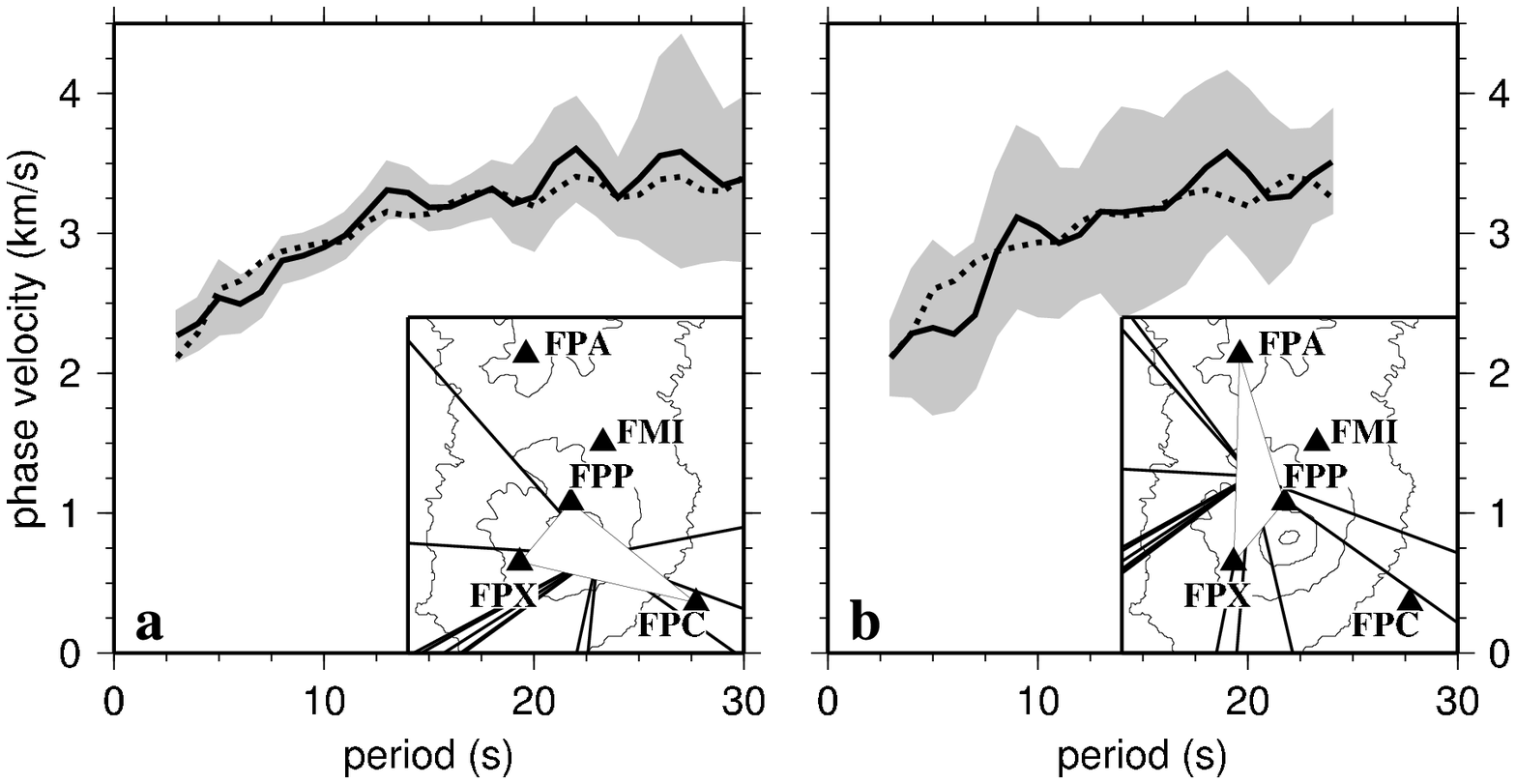} 
  \caption{}
 \label{fig5}
\end{figure}

\end{document}